\documentclass[article]{jss}

\usepackage{thumbpdf,lmodern}
\usepackage{tabularx}
\usepackage[toc,page]{appendix}
\usepackage{makecell}
\usepackage{soul}

\author{Zekai Wang\thanks{Corresponding Author: zwang@fairfield.edu; The first two authors developed the \pkg{reslife} package during their internship at UCB Pharma}\footnotemark[2]\\Fairfield \\ University\And
        Andrew Crawford \\The Georgia \\Institute of Technology \And 
        Ka Lok Lee\\UCB\\Pharma  \\ \And Lin Lu\\Fairfield\\University  \\ \And
        Srihari Jaganathan\\UCB\\Pharma}

\title{reslife: Residual Lifetime Analysis Tool in R}

\Plainauthor{Zekai Wang, Second Author} 
\Plaintitle{reslife: Residual Lifetime Analysis Tool in R} 
\Shorttitle{\pkg{reslife}: Residual Life in R} 

\Abstract{
   Mean residual lifetime is an important measure utilized in various fields, including pharmaceutical companies, manufacturing companies, and insurance companies for survival analysis. However, the computation of mean residual lifetime can be laborious and challenging. To address this issue, the \proglang{R} package \pkg{reslife} has been developed, which enables efficient calculation of mean residual lifetime based on closed-form solution in a user-friendly manner. \pkg{reslife} offers the capability to utilize either the results of a \pkg{flexsurv} regression or user-provided parameters to compute mean residual lifetime. Furthermore, there are options to return median and percentile residual lifetime. If the user chooses to use the outputs of a \pkg{flexsurv} regression, there is an option to input a data frame with unobserved data. In this article, we present \pkg{reslife}, explain its underlying mathematical principles, illustrate its functioning, and provide examples on how to utilize the package. The aim is to facilitate the use of mean residual lifetime, making it more accessible and efficient for practitioners in various disciplines, particularly those involved in survival analysis within the pharmaceutical industry. This package has been approved and available on CRAN: \url{https://cran.r-project.org/web/packages/reslife/index.html}
}
\Keywords{residual lifetime, survival analysis}
\Plainkeywords{keywords, comma-separated, not capitalized, Java} 

\begin{document}

\section{Motivation}

Survival analysis is a statistical approach widely used across numerous domains, including healthcare, economics, and the social sciences. A critical component of this analysis, particularly in the pharmaceutical industry, revolves around understanding the concept of residual lifetime, specifically the mean residual lifetime (MRL). MRL provides a measure of the expected remaining lifetime given that an individual has survived up to a certain time. This conditional expectation serves as a vital tool in assessing the efficacy of treatments, making prognoses, or predicting the reliability of systems. Despite its usefulness, the calculation of mean residual lifetime can be complex and time-intensive. Current approaches for MRL calculation often rely on numerical integration. However, these methods pose substantial challenges when dealing with large datasets, as the computational complexity increases considerably, making them impractical for real-world applications. Thus, there is a need for a computationally efficient method to calculate the MRL.

In order to address this gap, we present \code{reslife}, an R package specifically designed to simplify and accelerate the computation of the MRL. The unique aspect of our package is that, instead of relying on numerical integration, our package calculates the MRL based on closed-form solutions for the most common survival distributions, such as exponential, Weibull, gamma, generalized gamma distributions, and generalized F distribution. This offers a significant advantage in terms of speed and accuracy, making it especially suitable for large datasets.

\code{reslife} allows the use of outputs from user-provided parameters or from an R package called \code{flexsurv} \cite{jackson2016flexsurv}, which is for fully-parametric modeling of survival data, to calculate the mean residual lifetime. Moreover, it offers options to calculate the percentile residual lifetime. In addition, \code{reslife} also provides option to predict MRL for unobserved data.

The adoption of the reslife package demonstrates steady user engagement. According to cumulative download data from the RStudio CRAN mirror, the package has been downloaded approximately 3,793 times as of early 2025, with a consistent growth trend over time \citep{ipubDownloads, DataScienceMeta}. On average, 209 downloads per month have been recorded in recent periods. As of its current rank, reslife is positioned around 19,531st among CRAN packages based on total downloads \citep{cranlogsMonthly}.

\section{Review of Survival Modeling Packages in R}
Survival analysis is supported by a variety of R packages targeting specific analytics needs and methodological framework. Core packages such as \code{survival} \citep{therneau2015package} provides foundational capabilities for nonparametric, semiparametric, and parametric accelerated failure time survival modeling, while \code{flexsurv} \citep{jackson2016flexsurv} specializes in flexible parametric survival modeling, supporting Royston-Parmar spline model, generalized gamma and generalized F, and user-defined parametric distributions as well as expected survival times for both standard and complex survival models. Machine learning-based survival approaches, exemplified by \code{randomForestSRC} \citep{ishwaran2022package}, \code{ranger} \citep{wright2019package}, \code{survML}, and \code{survivalmodels}, offer powerful capabilities for capturing nonlinear relationships and interactions without the restrictive assumptions of traditional parametric methods. Bayesian survival modeling frameworks, including \code{rstanarm}, \code{brms} \citep{burkner2017brms}, \code{ROBSA} \citep{bartovs2022robsa}, and \code{psbcGroup} \citep{lee2017package}, provide probabilistic inference and robust uncertainty quantification. Joint modeling, effectively combining longitudinal and survival data, is supported by packages such as \code{JM} \citep{rizopoulos2010jm} and \code{JMbayes} \citep{rizopoulos2020package}, allowing dynamic and time-dependent survival predictions. 

Despite the breadth of functionalities available in existing survival analysis packages, a notable gap remains in the direct estimation and visualization of mean residual life (MRL), an intuitive and interpretable measure frequently used in clinical prognosis, maintenance scheduling, and reliability analysis. While \code{flexsurv} supports estimation of mean survival times for complex distributions, it does not directly compute or visualize MRL as a function over time, typically requiring manual numerical integration of survival functions. The newly introduced \code{reslife} package specifically addresses this gap by offering direct and user-friendly functions for computing and visualizing MRL. Its core advantage lies in explicitly providing closed-form MRL computations and visualizations for both common and complex survival distributions, including the generalized gamma and generalized F distributions. Although closed-form solutions for common distributions have previously been presented by \citet{poynor2019nonparametric}, the closed-form solutions for more complex distributions such as generalized gamma and generalized F have not been published before. The \code{reslife} package aims to enhance analytical flexibility, enabling practitioners and researchers to more accurately capture and interpret complex survival patterns observed in real-world data.

\section{MRL in General Parametric Survival Models}
\subsection{Properties of residual lifetime function}

In this subsection, we briefly discuss the fundamentals of residual lifetime function. First, the survival function of a positive random variable $T$ represents the probability that survival will extend beyond time $t$.

\begin{eqnarray}
S(t) = Pr(T > t) = 1 - F(t)
	\label{Eq: s}
\end{eqnarray}

where $F(t)$ is the cumulative distribution function (CDF). The mean residual lifetime (MRL) function calculates the anticipated remaining survival duration, conditioned on their survival up to a certain time, $x$. Specifically, MRL for a continuous variable can be represented by:

\begin{eqnarray}
MRL(x) = E(T-x|T>x) = \frac{\int_{x}^{\infty}(t-x)f(t)dt}{S(x)} = \frac{\int_{x}^{\infty}S(t)dt}{S(x)}
\label{Eq: mrl}
\end{eqnarray}

It is easy to show that when $x = 0$, the MRL function is equivalent to the expectation of $T$:

\begin{eqnarray}
MRL(0) = \frac{\int_{0}^{\infty}(t-0)f(t)dt}{S(0)} = \frac{\int_{0}^{\infty}(t-0)f(t)dt}{1} = E[T]
\label{Eq: mrl=0}
\end{eqnarray}

Alternatively, the MRL can be expressed in the following two forms: 

\begin{eqnarray}
MRL(x) = \frac{\int_{x}^{\infty}(t-x)f(t)dt}{S(x)} = \frac{\int_{x}^{\infty}tf(t)dt}{S(x)} - x
\label{Eq: m(x)2}
\end{eqnarray}

\begin{eqnarray}
MRL(x) = \frac{\int_{x}^{\infty}(t-x)f(t)dt}{S(x)} = \frac{E[T] - \int_{0}^{x}S(t)dt}{S(x)}
\label{Eq: m(x)3}
\end{eqnarray}

\begin{table}[ht]
\centering
\begin{tabular}{|c|c|c|}
\hline
Name & parameters & Argument  \\
\hline
Exponential & \code{rate} & \code{"exponential"} \\
Weibull & \code{shape}, \code{scale} & \code{"weibull"} \\
Gamma  & \code{shape}, \code{rate}  & \code{"gamma"} \\
Gompertz & \code{shape}, \code{rate}  & \code{"gompertz"} \\
Log-normal & \code{meanlog}, \code{sdlog} & \code{"lnorm"} \\
Log-logistic & \code{shape}, \code{scale} & \code{"llogis"} \\
Generalized gamma \cite{stacy1962generalization}& \code{shape}, \code{scale}, \code{k} & \code{"gengamma.orig"} \\
Generalized gamma \cite{prentice1975discrimination}&  \code{mu}, \code{sigma}, \code{Q} & \code{"gengamma"} \\
Generalized F \cite{prentice1975discrimination}&  \code{mu}, \code{sigma}, \code{s1},\code{s2} & \code{"genf.orig"} \\
Generalized F \cite{prentice1975discrimination}&  \code{mu}, \code{sigma}, \code{Q},\code{P} & \code{"genf"} \\
\hline
\end{tabular}
\caption{Built-in parametric survival distributions in reslife}
\label{tab:built-in}
\end{table}

We utilize the aforementioned properties of the MRL when calculating the closed-form MRL function for common parametric survival distributions in subsequent sections. For an in-depth exploration and proof of the MRL function's properties, see \cite{poynor2010bayesian}. 

The percentile lifetime function (PLF) is the time at which a certain percentage of the population has failed. A special case of percentile lifetime function is a median residual lifetime (i.e., the time at which 50\% of the population has failed.). Calculating the Percentile Residual Lifetime (PRL) is relatively straightforward given the survival function of parametric distribution. Specifically, for $0<\alpha < 1$, the $\alpha$-percentile residual lifetime function $q_{\alpha}(x)$ is defined as:

\begin{eqnarray}
q_{\alpha}(x) = F^{-1}(1 - (1-\alpha)S(x)) - x
\label{Eq: plf}
\end{eqnarray}

where $F^{-1}(.)$ is the inverse function of CDF. When $\alpha = 0.5$, the percentile residual lifetime function is the median residual lifetime function:

\begin{eqnarray}
MedianRL(x) = q_{0.5}(x) = F^{-1}(1 - \frac{1}{2}S(x)) - x
\label{Eq: plf2}
\end{eqnarray}

On the other hand, the calculation of the mean residual lifetime (MRL) can be quite difficult due to the need to integrate complex probability density function or survival function of parametric distributions (See Eq.\ref{Eq: mrl} and Eq.\ref{Eq: m(x)2}). In the upcoming subsection, we will introduce the closed-form solution for the MRL under the most commonly used parametric survival distributions.

\subsection{MRL for parametric survival distributions}
The parametric distributions that are currently incorporated into our package for the computation of residual lifetime are listed in Table 1. All the parametric distributions mentioned in Table 1 have closed-form mean residual lifetime (MRL), the details of which can be referenced in Table 2. Note that $S(t)$ represents the survival function at $t$; $\Gamma(\cdot)$ represents gamma function; $\Gamma(\cdot,\cdot)$ means incomplete gamma function\footnote{definition given in equation (\ref{upper_incgamma}) of Appendix}; $\Phi(\cdot)$ means the CDF of normal distribution. Complete proofs and notations for the Exponential, Weibull, Gamma, Gompertz, Log-normal, and Log-logistic distributions can be referenced in \cite{poynor2010bayesian}. Detailed proofs for the two Generalized Gamma distributions and two Generalized F distributions are available in the Appendix, and this is the first time the closed-form expressions of MRL for these distributions are presented to the best of our knowledge.

\begin{table}
\renewcommand{\arraystretch}{4}
\setlength{\tabcolsep}{8 pt}
\centering
\begin{tabular}{c| c c}
\hline
Name & parameters & MRL  \\
\hline
\makecell{Exponential\\ (\code{exp})}& \code{rate}=$\lambda$ & $\lambda$ \\
\makecell{Weibull\\ (\code{weibull})} & \code{shape}=$\alpha$, \code{scale} = $\lambda$ & \makecell{$\frac{\lambda}{\alpha}\Gamma(\frac{1}{\alpha})S(x^\alpha)/S(x)$}\\
\makecell{Gamma\\ (\code{gamma})}  & \code{shape}=$\alpha$, \code{rate} = $\lambda$  & $x^{\alpha}\texttt{exp}(-\frac{x}{\lambda})/[\lambda^{\alpha-1}\Gamma(\alpha)S(x)$] \\
\makecell{Gompertz\\ (\code{gompertz})} & \code{shape}=$\alpha$, \code{rate} = $\lambda$  & \makecell{$\texttt{exp}(\frac{\lambda}{\alpha}e^{\alpha x})(\frac{1}{\alpha})\Gamma(0,\frac{\lambda}{\alpha}e^{\alpha x})$} \\
\makecell{Log-normal\\ (\code{lnorm})} & \code{meanlog}=$\mu$, \code{sdlog}=$\sigma$ & $\texttt{exp}(\mu + \frac{\sigma^2}{2})[1-\Phi(\frac{ln(x)-(\mu+\sigma^2)}{\sigma})]/S(x) - x$ \\
\makecell{Log-logistic\\ (\code{llogis})} & \code{shape}=$\alpha$, \code{scale}= $\lambda$ & \makecell{$\frac{\lambda}{\alpha}\Gamma(1-\frac{1}{\alpha})\Gamma(\frac{1}{\alpha})S(z(x);1-\frac{1}{\alpha},\frac{1}{\alpha})(1 + \frac{x}{\lambda})^\alpha$ \\(where $z(x)=(\frac{x}{\lambda})^{\alpha}/(1 + (\frac{x}{\lambda})^\alpha)$)} \\
\makecell{Generalized gamma\\ \cite{stacy1962generalization}\\ (\code{gengamma.orig})}& \makecell{\code{shape} = b, \code{scale} = a, \\\code{k} = k }& \makecell{See Appendix A }\\
\makecell{Generalized gamma\\ \cite{prentice1974loggamma}\\ (\code{gengamma})}&  \makecell{\code{mu}=$\mu$, \code{sigma}=$\sigma$, \\ \code{Q} = $Q$} & \makecell{See Appendix B} \\
\makecell{Generalized F\\ \cite{prentice1975discrimination}\\ (\code{genf.orig})}&  \makecell{\code{mu}=$\mu$, \code{sigma}=$\sigma$, \\ \code{s1} = s1, \code{s2} = s2} & See Appendix C \\
\makecell{Generalized F\\ \cite{prentice1975discrimination}\\ (\code{genf})}&  \makecell{\code{mu}=$\mu$, \code{sigma}=$\sigma$, \\ \code{Q} = $Q$, \code{P} = $P$} & See Appendix D \\
\hline
\end{tabular}
\caption{Closed-form MRLs for built-in parametric survival distributions}
\label{tab:mrl_closeform}
\end{table}

\section{Implementations}
In this section, we will demonstrate the \pkg{reslife} package with reproducible examples and show how to use the \pkg{reslife} with the real-world data. 
The \pkg{reslife} package contains two functions, \code{reslifefsr} and \code{residlife}. The \code{reslifefsr} function relies on the output of a \code{flexsurvreg} object to compute MRL, while \code{residlife} accepts user-specified distribution parameters directly. Users may directly use the parameters of \code{residlife} to vectorize the computation. An overview and example of both are provided.  
\subsection{reslifefsr overview}
The function \code{reslifefsr} is for use after the user creates a \code{flexsurvreg} output object, and it has the form

\code{reslifefsr(obj, life, p = .5, type = 'mean', newdata = data.frame())
}

The first argument, \code{obj}, is the name of a  \code{flexsurvreg} object. All necessary information about the distribution and its parameters are extracted from this object. The second argument, \code{life}, is a positive scalar representing the lifetime elapsed. Both of these arguments must be provided for the function to run. The remaining arguments are optional.  \code{p} is a value between 0 and 1 specifying the percentile at which residual life will evaluated, should the user desire to see percentile residual lifetime. The default value is \code{".5"}, representing the median residual lifetime. 

\code{type} tells the function which residual lifetime to output. The options are \code{"mean"}, \code{"median"}, \code{"percentile"}, or \code{"all"}, with the default being \code{"mean"}. 

\code{newdata} allows the user to provide new data to the function. This data must be provided in a \code{"data.frame"} object. The default option is a blank \code{data.frame}. The columns names of \code{newdata} must, at a minimum, contain the regressors used in the \code{flexsurvreg} object. Columns other than those used in the \code{flexsurvreg} are allowed, but will be ignored by the function. If no new data is provided, the function will use the data from the \code{flexsurvreg} object.

\subsection{reslifefsr demo}

To demonstrate how to use \code{reslifefsr}, we go through the following example. First, we show how the function's output compares to the result of numerical integrating. Specifically, we compute the MRL using the package and compare it to a direct numerical integration for the generalized gamma case. We similarly validate the MRL formulas for generalized F distributions by comparing to integration. We start by creating a \code{flexsurvreg} object based on the \code{"bc"} dataset, which is built into \code{flexsurv}. Let's look at the bc dataset.

The bc dataset, available in the R package \code{flexsurv}, contains survival data for breast cancer patients, including the variable \textit{recyrs} representing time to recurrence or censoring (in years), and the event indicator \textit{censrec}. Patients are stratified into prognostic risk groups based on a regression model developed by Sauerbrei and Royston (1999), with the \textit{group} variable indicating their assigned level of recurrence risk. This dataset serves as a practical example for illustrating survival analysis techniques such as model fitting, group comparisons, and estimation of residual life.

\begin{CodeChunk}
\begin{CodeInput}
R> library(reslife)
R> library(flexsurv)
R> data(bc)
R> force(bc)
\end{CodeInput}
\begin{CodeOutput}
    censrec rectime  group     recyrs
1         0    1342   Good 3.67671233
2         0    1578   Good 4.32328767
3         0    1760   Good 4.82191781   
\end{CodeOutput}
\end{CodeChunk}

\code{recyrs} is our variable of interest in this data set. For the first example, the covariate is just the intercept term. We will set the \code{life} value to 1 for this example. 

\begin{CodeChunk}
\begin{CodeInput}
R> fs1 <- flexsurvreg(Surv(recyrs, censrec) ~ 1, data = 
+  bc, dist = "gengamma.orig")
R> reslifefsr(obj = fs1, life = 1, type = 'mean')
\end{CodeInput}
\begin{CodeOutput}
[1] 7.193907
\end{CodeOutput}
\end{CodeChunk}
For someone who has survived 1 year, we can expect them to survive another 7.193907 years. 

Now we can compute the mean residual lifetime using numerical integration (i.e., use the built-in numerical integration function, \code{integrate}) to verify it matches the output of the \code{reslifefsr} function. 

\begin{CodeChunk}
\begin{CodeInput}
R> b <- exp(as.numeric(fs1$coefficients[1]))
R> a <- exp(as.numeric(fs1$coefficients[2]))
R> k <- exp(as.numeric(fs1$coefficients[3]))
R> f <- function(x) {
+    return(pgengamma.orig(x, shape = b, scale = a, k = k,
+    lower.tail = FALSE))
+   }
R> integrate(f, 1, Inf)$value/pgengamma.orig(1, shape = b,
+  scale = a, k = k, lower.tail = FALSE)
\end{CodeInput}
\begin{CodeOutput}
[1] 7.193907
\end{CodeOutput}
\end{CodeChunk}

The two values are the same. 
Now we can do a demonstration with multiple regressors. First, the \code{"bc"} data set needs to be modified to include a continous variable called age. Then, a new \code{flexsurvreg} object is created. We use the \code{weibull} distribution for this second \code{flexsurvreg}.

\begin{CodeChunk}
\begin{CodeInput}
R> newbc <- bc
R> newbc$age <- rnorm(dim(bc)[1], mean = 65-
+  scale(newbc$recyrs, scale=FALSE),sd = 5)
R> fsr2 <-  flexsurvreg(Surv(recyrs, censrec) ~ group+age,
+  data=newbc, dist = 'weibull')

\end{CodeInput}
\end{CodeChunk}
We can now use the \code{reslifefsr} function to find the mean residual lifetime. We will use a \code{life} value of 4 and will limit the responses to 12 in order to save space. The full result has 686 values, one for each observation in the data set. 

\begin{CodeChunk}
\begin{CodeInput}
R> reslifefsr(obj = fsr2, life= 4)
\end{CodeInput}
\begin{CodeOutput}
[1]  5.2727227  6.8176956  5.8068081  6.1437970
[5]  6.2348949  7.8047753  4.1965165  6.7279331
[9] 10.7687335 10.1025738  9.6337556  9.7637072
\end{CodeOutput}
\end{CodeChunk}

We can repeat that for the median residual lifetime. 
\begin{CodeChunk}
\begin{CodeInput}
R> reslifefsr(obj = fsr2,life= 4, type = 'median')
\end{CodeInput}
\begin{CodeOutput}
[1]  4.1536887  5.4356404  4.5955187  4.8750693
[5]  4.9507329  6.2597970  3.2690174  5.3608650
[9]  8.7496077  8.1885035  7.7940740  7.9033654
\end{CodeOutput}
\end{CodeChunk}
And finally for all three types of residual lifetime. \code{p} is set to \code{".8"}, so we are looking at the 80th percentile. Since \code{"all"} is specified for \code{type}, the output is a dataframe with each column a different measure of residual life. 
\begin{CodeChunk}
\begin{CodeInput}
R> reslifefsr(obj = fsr2, life= 4, p = .8, type = 'all')
\end{CodeInput}
\begin{CodeOutput}
         mean    median percentile
1    5.272723  4.153689   8.494733
2    6.817696  5.435640  10.945314
3    5.806808  4.595519   9.343409
4    6.143797  4.875069   9.878029
5    6.234895  4.950733  10.022444
6    7.804775  6.259797  12.504669
7    4.196517  3.269017   6.778863
8    6.727933  5.360865  10.803287
9   10.768734  8.749608  17.164975
10  10.102574  8.188503  16.119987
11   9.633756  7.794074  15.383793
12   9.763707  7.903365  15.587926
\end{CodeOutput}
\end{CodeChunk}

Another capability of \code{reslifefsr} is its ability to work with new, user-supplied data. We can create an example data frame and supply that to the function to generate predictions. 
\begin{CodeChunk}
\begin{CodeInput}
R> group <- c("Medium", 'Good', "Poor")
R> age <- c(43, 35, 39)
R> newdata <- data.frame(age, group)
R> reslifefsr(obj = fsr2,life= 4,p = .6, type = 'all', newdata= newdata)

\end{CodeInput}
\begin{CodeOutput}
       mean    median percentile
1 10.322444  8.373620  10.519207
2 27.539287 22.982792  28.406780
3  6.235569  4.951293   6.283933

\end{CodeOutput}
\end{CodeChunk}

The function is able to handle some variation in the input data frame. Changing the order of the columns does not have an impact on the results, nor does having more columns than needed. The only requirements are all regressors in the \code{flexsurvreg} object must be present in the data frame, and if the variables are factors, no factor levels outside the scope of the original data set are allowed.

\begin{CodeChunk}
\begin{CodeInput}
R> newdata <- data.frame(group, age)
R> reslifefsr(obj = fsr2,life= 4,p = .6, type = 'all', newdata= newdata)
\end{CodeInput}
\begin{CodeOutput}
       mean    median percentile
1 10.322444  8.373620  10.519207
2 27.539287 22.982792  28.406780
3  6.235569  4.951293   6.283933
\end{CodeOutput}
\end{CodeChunk}

\begin{CodeChunk}
\begin{CodeInput}
R> extra = c(100, 100, 100)
R> newdata2 = data.frame(age, group, extra)
R> reslifefsr(obj = fsr2,life= 4,p = .6, type = 'all', newdata= newdata2)
\end{CodeInput}
\begin{CodeOutput}
       mean    median percentile
1 10.322444  8.373620  10.519207
2 27.539287 22.982792  28.406780
3  6.235569  4.951293   6.283933
\end{CodeOutput}
\end{CodeChunk}

\begin{CodeChunk}
\begin{CodeInput}
R> newdata3 = data.frame(group)
R> reslifefsr(obj = fsr2, life = 4,p = .6, type = 'all', newdata= newdata3)
\end{CodeInput}
\begin{CodeOutput}
Error in weibull_mlr(obj, life, p, type, newdata) : 
 fsoutput$covdata$covnames == colnames(newdata) are not all TRUE
\end{CodeOutput}
\end{CodeChunk}
The function returns an error since fewer columns are 
provided than what is required. 
\code{flexsurvreg}. 

Providing an incorrect factor level also throws an error. The levels for \code{group} are 
\code{Good}, \code{Medium}, and \code{Poor}, so including \code{Terrible} is incorrect. 

\begin{CodeChunk}
\begin{CodeInput}
R> group = c("Medium", 'Good', "Terrible")
R> newdata = data.frame(age, group)
R> reslifefsr(obj = fsr2,life= 4,p = .6, type = 'all', newdata= newdata)
\end{CodeInput}
\begin{CodeOutput}
[1] "Incorrect Level Entered"
Error in weibull_mlr(obj, life, p, type, newdata) : error
\end{CodeOutput}
\end{CodeChunk}

It is also possible to plot the mean residual lifetime for a set of values from a 'flexsurv' object.

We take the first line from newbc and create a blank vector of length 10. We then loop the function to 
create the MRLs for the first 10 integers, which gives us the expected survival time given the patient has 
survived y years. 

We can then plot these values. Figure 1 shows the mean residual life over time. 

\begin{CodeChunk}
\begin{CodeInput}
R> newdata4 = newbc[1,]
R> mrl = rep(0,10)
R> for (i in 1:10){
R> mrl[i] = reslifefsr(obj = fsr2,life= i,p = .6, type = 'mean', newdata= newdata4)
R> }
R> plot(seq(1,10),mrl,xlab="Survival Time",
+  main = 'MRL over Time', ylab = "MRL",type="o",ylim=c(0,7))

\end{CodeInput}
\end{CodeChunk}

\begin{figure}
	\begin{center}
		\includegraphics{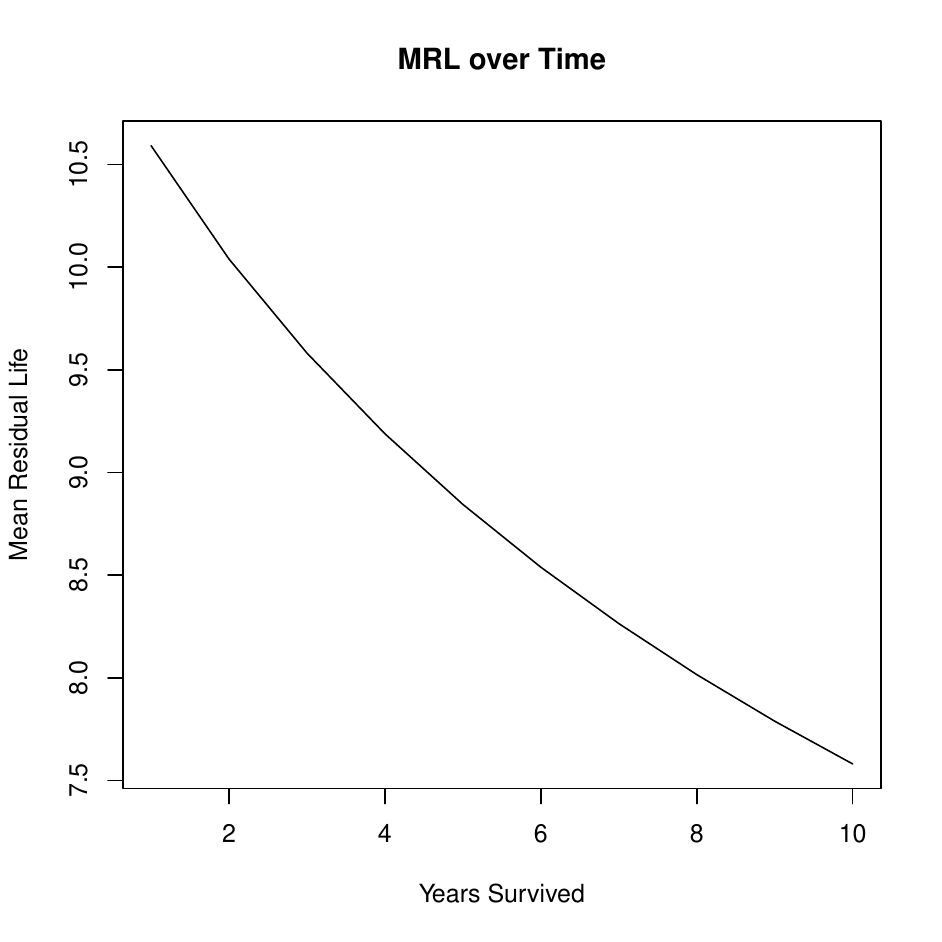}
		\caption{Mean residual life over years survived for a breast cancer patient using the \code{reslifefsr} function.}
		\label{Fig:plot}
	\end{center}      	
\end{figure}

\subsection{residlife overview}

The function \code{residlife} is for user-inputted parameters and it has the form:

\code{residlife(values, distribution, parameters, p = .5, type = 'mean')
}

The first argument, \code{values}, are the values of \code{life} the user wants to see a residual lifetime value for. \code{values} can either be a scalar or a vector. The function will return all the specified residual lifetime for all values in the vector. 

\code{distribution} specifies the distribution the user would like to see. The argument can be any one of \code{"weibull"}, \code{"gompertz"}, \code{"gamma"}, \code{"gengamma"}, \code{"gengamma.orig"}, \code{"genf"}, \code{"genf.orig"}, \code{"exponential"}, \code{"lnorm"}, or \code{"llogis"}. 

\code{parameters} defines the parameters for the chosen distribution. These parameters must be inputted as a vector, with the name of each parameter correctly specified. These parameters must be in order, as well. For a distribution such as \code{"gamma"}, where there are multiple parameterizations, the function has some flexibility. Both \code{(shape = , scale = )} and \code{(shape = , rate = )} are acceptable for the \code{gamma} distribution. 

The remaining arguments are optional, and follow the same format as in the \code{reslifefsr} function. \code{p} is a value between 0 and 1 which specifies the percentile at which residual lifetime is calculated. The default value is \code{".5"}. 

\code{type} specifies which residual lifetime to output. The options are \code{"mean"}, \code{"median"}, \code{"percentile"}, or \code{"all"}, with the default being \code{"mean"}. 

A major difference between \code{reslifefsr} and \code{residlife} is that \code{residlife} is univariate, as the exact values of the parameters must be supplied. 

\subsection{residlife demo}

We will now give a demonstration of how the \code{residlife} function works. 
For the examples in this vignette, we will use the output from a \code{flexsurvreg} using the \code{newbc} dataset.

\begin{CodeChunk}
\begin{CodeInput}
R> fsr3 = flexsurvreg(Surv(recyrs, censrec) ~1, 
+  data=newbc, dist = 'weibull')
R> fsr3$res    
\end{CodeInput}
\begin{CodeOutput}
           est     L95%     U95%         se
shape 1.271519 1.153372 1.401770 0.06326773
scale 6.191377 5.604203 6.840071 0.31475730   
\end{CodeOutput}
\end{CodeChunk}
We use a sequence of values from 1 to 10, counting up by .5, for these examples
\begin{CodeChunk}
\begin{CodeInput}
R> residlife(values = seq(1,10,.5), distribution = 'weibull', 
+  parameters= c(shape= 1.272, scale = 6.191))
\end{CodeInput}
\begin{CodeOutput}
[1] 5.280618 5.125825 4.994025 4.878801 4.776287 4.683907 4.599837 4.522725 4.451537 
[10] 4.385457 4.323832 4.266128 4.211904 4.160787 4.112464 4.066666 4.023161
[19] 3.981746 3.942246
\end{CodeOutput}
\end{CodeChunk}

The parameter names inputted in the \code{parameter} argument must be correct for the specified distribution. Misaligned distribution and parameter names will return an error. 

\begin{CodeChunk}
\begin{CodeInput}
R> residlife(values = seq(1,10,.5), distribution= 'weibull',
+  parameters= c(shape= 1.272, not_scale = 6.191))
\end{CodeInput}
\begin{CodeOutput}
[1] "incorrect parameters entered. Parameters for weibull are shape and scale"
Error in residlife(values = seq(1, 10, 0.5), distribution = "weibull", : error
\end{CodeOutput}
\end{CodeChunk}

As mentioned above, the function can handle different parameterizations for distributions with multiple options. 
Below uses a \code{gamma} distribution with \code{shape} and \code{scale} parameters. 

\begin{CodeChunk}
\begin{CodeInput}
R> residlife(values = seq(1,10,.5), distribution= 'gamma', 
+  parameters= c(shape= 1.272, scale = 6.191))
\end{CodeInput}
\begin{CodeOutput}
[1] 7.498217 7.389908 7.302911 7.230487 7.168736 7.115159 7.068047 7.026172 6.988622 
[10] 6.954700 6.923859 6.895665 6.869766 6.845874 6.823747 6.803186 6.784020 
[19] 6.766103 6.74931
\end{CodeOutput}
\end{CodeChunk}

The result is the same if we use \code{shape} and \code{rate} parameters. 

\begin{CodeChunk}
\begin{CodeInput}
R> residlife(values = seq(1,10,.5), distribution= 'gamma',
+  parameters= c(shape= 1.272, rate = 1/6.191))
\end{CodeInput}
\begin{CodeOutput}
[1] 7.498217 7.389908 7.302911 7.230487 7.168736 7.115159 7.068047 7.026172 6.988622 
[10] 6.954700 6.923859 6.895665 6.869766 6.845874 6.823747 6.803186 6.784020 
[19] 6.766103 6.74931
\end{CodeOutput}
\end{CodeChunk}

Much like the \code{reslifefsr} function, \code{residlife} has the ability to show \code{mean}, \code{median}, \code{percentile}, or\code{all} three types of residual life. Below is an example of the function with \code{type} set to \code{'all'}.

\begin{CodeChunk}
\begin{CodeInput}
R> residlife(values = seq(1,10,.5), distribution= 'weibull', 
+  parameters= c(shape= 1.272, scale = 6.191), p = .7, type = 'all')
\end{CodeInput}
\begin{CodeOutput}
      mean   median percentile
1  5.280618 4.151524   6.619995
2  5.125825 3.988283   6.423757
3  4.994025 3.851217   6.253251
4  4.878801 3.733262   6.102230
5  4.776287 3.629997   5.966703
6  4.683907 3.538406   5.843883
7  4.599837 3.456318   5.731711
8  4.522725 3.382114   5.628606
9  4.451537 3.314545   5.533322
10 4.385457 3.252635   5.444853
11 4.323832 3.195598   5.362376
12 4.266128 3.142799   5.285206
13 4.211904 3.093714   5.212767
14 4.160787 3.047908   5.144570
15 4.112464 3.005015   5.080196
16 4.066666 2.964725   5.019284
17 4.023161 2.926773   4.961518
18 3.981746 2.890929   4.906624
19 3.942246 2.856997   4.854361
\end{CodeOutput}
\end{CodeChunk}

There are instances where the function will return a \code{NaN}, but that has to do more with the values inputted than the function itself. Certain combinations of life values and distribution parameters are not calculable. We will show one example below. The \code{shape} parameter has been changed to 2.9, the \code{scale} to 2.2 and the \code{values} is a vector from 15 to 30.  
\begin{CodeChunk}
\begin{CodeInput}
R> residlife(values = seq(15,30), distribution= 'weibull', 
+  parameters= c(shape= 2.9, scale = 2.2), p = .7, type = 'all')
\end{CodeInput}
\begin{CodeOutput}
         mean      median percentile
1  0.01972296 0.013693147 0.02376936
2  0.01745426 0.012114707 0.02103170
3  0.01556040 0.010797874 0.01874721
4  0.01396275 0.009687539 0.01682059
5  0.01260229 0.008742418 0.01518039
6  0.01143404 0.007931086 0.01377220
7  0.01042322 0.007229272 0.01255397
8         NaN         Inf        Inf
9         NaN         Inf        Inf
10        NaN         Inf        Inf
11        NaN         Inf        Inf
12        NaN         Inf        Inf
13        NaN         Inf        Inf
14        NaN         Inf        Inf
15        NaN         Inf        Inf
16        NaN         Inf        Inf
\end{CodeOutput}
\end{CodeChunk}

The life values after 7 are not feasible, with the mean function returning a \code{NaN} value and the median and percentile residual life returning a \code{Inf} value. 

The outputs of this function can then easily be taken and used in a plot or any other type of
analysis tool. 

\begin{figure}
	\begin{center}
	\includegraphics{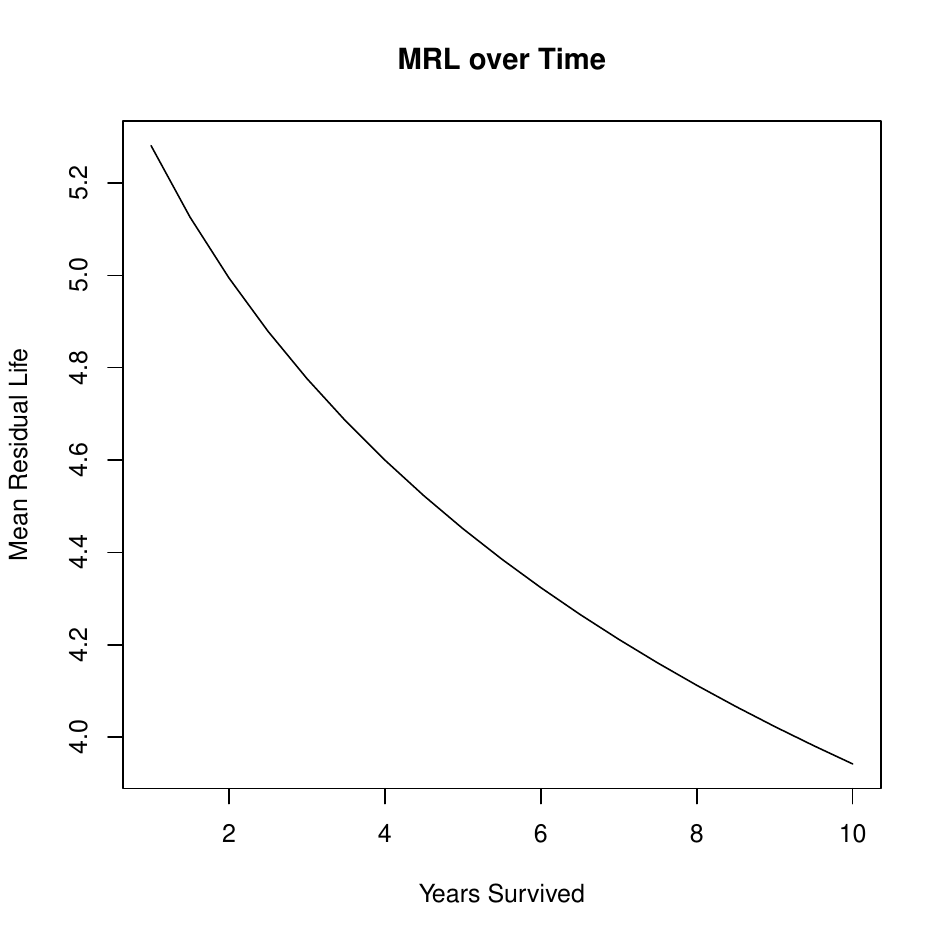}
		\caption{Mean residual life over years survived for a group of breast cancer patients, computed using the \code{residlife} function with a fitted Weibull distribution.}
		\label{Fig:plot2}
	\end{center}      	
\end{figure}

Let's look at an example of such a plot in Figure \ref{Fig:plot2}, which shows the mean residual life over time. 

Say we want to plot the mean residual lifetime over time. First, we would create a vector for time, 
run it through the residlife() function with type = 'mean', and plot the output against time.

After extracting the parameters, we can run residlife() and generate an output vector. 
Note that since residlife() can take a vector input, we don't have to use a loop.  


\begin{CodeChunk}
\begin{CodeInput}
R> time= seq(1,10,.5)
R> life = residlife(values = time, distribution = 'weibull',
+  parameters=  c(shape= 1.272, scale = 6.191))
R> plot(time, life,xlab="Years Survived",ylab = "Mean Residual Life"
+  main = 'MRL over Time, type='l')


\end{CodeInput}
\begin{CodeOutput}
    
\end{CodeOutput}
\end{CodeChunk}

\section{Conclusion}

The \pkg{reslife} package serves as a valuable tool for calculating mean residual lifetime and the related percentile residual lifetime, both are crucial metrics in survival analysis across multiple fields. To the best of our knowledge, there is no package in \code{R} that can do this. By simplifying the calculation of mean residual lifetime, the package empowers researchers and practitioners to gain valuable insights more efficiently, contributing to advancements in survival analysis in various disciplines, including the pharmaceutical industry.

\begin{appendix}
\section{MRL Derivation for Original Generalized Gamma}
The PDF of the original generalized gamma (GG) distribution from \cite{stacy1962generalization} is:

\begin{eqnarray}
\label{pdf}
f(t;a,b,k) = \frac{bt^{bk-1}}{\Gamma(k)a^{bk}}\texttt{exp}((-\frac{t}{a})^b)
\end{eqnarray}

where $a$,$b$,$k$ are parameters of GG, $\Gamma(k)$ is Gamma function.

We represent Mean Residual Lifetime (MRL) function using:

\begin{eqnarray}
\label{mrl1}
MRL(x) = \frac{\int_{x}^{\infty}tf(t)dt}{S(x)} - x
\end{eqnarray}

The numerator of equation \ref{mrl1} can be written as:

\begin{eqnarray}
\label{numerator_gg}
\int_{x}^{\infty}tf(t) &=& \int_{x}^{\infty}t\frac{bt^{bk-1}}{\Gamma(k)a^{bk}}\texttt{exp}((-\frac{t}{a})^b)dt\\ \nonumber
&=& \frac{b}{\Gamma(k)}\int_{x}^{\infty}(\frac{t}{a})^{bk}\texttt{exp}((-\frac{t}{a})^b)dt
\end{eqnarray}

Let $u = (\frac{t}{a})^b$, then we can obtain $du = \frac{b}{a}(\frac{t}{a})^{b-1}dt$, and $dt = \frac{a}{b}(u)^{\frac{1-b}{b}}du$

Thus, the new limit for $u$ become $u = (\frac{x}{a})^b$ when $t=x$ and $u \rightarrow \infty$, when $t \rightarrow \infty$.

Make a substitution to simplify the integral in equation \ref{numerator_gg}:

\begin{eqnarray}
\label{simple1}
\int_{x}^{\infty}tf(t)dt &=& \frac{b}{\Gamma(k)}\int_{x}^{\infty}(\frac{t}{a})^{bk}\texttt{exp}((-\frac{t}{a})^b)dt \\ \nonumber
&=& \frac{b}{\Gamma(k)} \int_{(\frac{x}{a})^b}^{\infty}u^k\texttt{exp}(-u) \frac{a}{b} u^{\frac{1-b}{b}}du \\ \nonumber
&=& \frac{a}{\Gamma(k)} \int_{(\frac{x}{a})^b}^{\infty}u^{k + \frac{1}{b} - 1}\texttt{exp}(-u)du
\end{eqnarray}

The integral part of equation \ref{simple1} is known as the upper incomplete gamma function, 

\begin{eqnarray}
\label{upper_incgamma}
\Gamma(a,x) = \int_{x}^{\infty}t^{a-1}\texttt{exp}(-t)dt
\end{eqnarray}

which become $\Gamma(k+\frac{1}{b},(\frac{x}{a})^b)$.

Thus,

\begin{eqnarray}
\label{finalpart}
\int_{x}^{\infty}tf(t)dt &=& a \frac{\Gamma(k+\frac{1}{b},(\frac{x}{a})^b)}{\Gamma(k)}
\end{eqnarray}

Since the survival function of GG can be represented as:

\begin{eqnarray}
\label{sf}
S(x) = \frac{\Gamma(k,(\frac{x}{a})^{b})}{\Gamma(k)}
\end{eqnarray}

We can express MRL for original GG (i.e., \code{gengemma.orig}) as:

\begin{eqnarray}
\label{final_ggorig}
MRL(x) &=& a\frac{\Gamma(k+\frac{1}{b},(\frac{x}{a})^b)}{\Gamma(k,(\frac{x}{a})^{b})} - x
\end{eqnarray}

\section{MRL Derivation for Prentice Generalized Gamma}
The PDF of the generalized gamma (GG) distribution from \cite{prentice1974loggamma} using \cite{cox2007gamma}'s expression is:
\begin{eqnarray}
\label{pdf2}
f(t;\mu, \sigma, Q) = \frac{|Q|}{\sigma t \Gamma(Q^{-2})}[Q^{-2}(e^{-\mu}t)^{\frac{Q}{\sigma}}]^{Q^{-2}}\texttt{exp}[-Q^{-2}(e^{-\mu}t)^{\frac{Q}{\sigma}}] 
\end{eqnarray}
where $\mu$,$\sigma$,$Q$ are parameters of GG, $\Gamma(\lambda^{-2})$ is Gamma function.

\cite{jackson2016flexsurv} provides an identity between Prentice Generalized Gamma and the original Generalized Gamma from \cite{stacy1962generalization} for $Q > 0$: 
\begin{CodeChunk}
\begin{CodeInput}
R> dgengamma.orig(x, shape=abs(Q)/(sigma), 
   scale=exp(mu-(log(Q^(-2))*sigma)/(abs(Q))),
   k=1/(Q^2)) == dgengamma(x, mu=mu, sigma=sigma, Q=abs(Q))
R> dgengamma.orig(x, shape=b, scale=a, k) == 
   dgengamma(x, mu=log(a) + log(k)/b, sigma=1/(b*sqrt(k)), Q=1/sqrt(k))
\end{CodeInput}
\end{CodeChunk}
Please note that machine epsilon may falsely suggest the above identities do not hold.\par 

For $Q < 0$, we can use this MRL representation:
\begin{eqnarray}
\label{mrlgg}
MRL(x) = \frac{\int_{x}^{\infty}S(t)dt}{S(x)}
\end{eqnarray}

\cite{cox2007gamma} provides the corresponding survival function in terms of lower incomplete gamma function, $\gamma(\cdot,\cdot)$:
\begin{eqnarray}
\label{Sgg}
S_{GG}(t) = \frac{\gamma(Q^{-2},Q^{-2}(e^{-\beta}t)^{\frac{Q}{\sigma}})}{\Gamma(Q^{-2})} = \frac{\int_{0}^{Q^{-2}(e^{-\beta}t)^{\frac{Q}{\sigma}}}z^{Q^{-2}-1}e^{-z}dz}{\Gamma(Q^{-2})}
\end{eqnarray}

The numerator of equation \ref{mrlgg} is what to focus on:

\begin{eqnarray}
\label{numerator_mrlgg}
\int_{x}^{\infty}\int_{0}^{Q^{-2}(e^{-\mu}t)^{\frac{Q}{\sigma}}}z^{Q^{-2}-1}e^{-z}dzdt
\end{eqnarray}

Changing the order of integration, we get:

\begin{eqnarray}
\int_{0}^{Q^{-2}(e^{-\mu}(Q^{2}x)^{\frac{Q}{\sigma}}}[\int_{x}^{e^{\mu}(Q^{-2}z)^{\frac{\sigma}Q}}\texttt{1}dt]z^{Q^{-2}-1}e^{-z}dz
\end{eqnarray}

For the ease of exposition, we let 
$A = Q^{-2}(e^{-\mu}(Q^{2}x)^{\frac{Q}{\sigma}}$.

\begin{eqnarray}
=\int_{0}^{A}(e^{\mu}(Q^{-2}z)^{\frac{\sigma}{Q}}-x)z^{Q^{-2}-1}e^{-z}dz 
\end{eqnarray}
\begin{eqnarray}
=e^{\mu}(Q^{-2})^{\frac{\sigma}{Q}}\int_{0}^{A}z^{\frac{\sigma}{Q}+Q^{-2}-1}e^{-z}dz-x\int_{0}^{A}z^{Q^{-2}-1}e^{-z}dz
\end{eqnarray}

Bringing back the denominator $S_{GG}(x)$, $MRL(x)$ for Prentice Generalized Gamma (i.e., \code{gengamma}) when $Q<0$ is
\begin{eqnarray}
\label{MRL_ggpren}
\frac{e^{\mu}(Q^{-2})^{\frac{\sigma}{Q}}\gamma(\frac{\sigma}{Q}+Q^{-2},A)-x\gamma(Q^{-2},A)}{\gamma(Q^{-2},A)}
\end{eqnarray}

For practical implementation, we use standard mathematical functions, such as the incomplete gamma function, provided by common computational libraries to evaluate these expressions.

\section{MRL Derivation for Original Generalized F Distribution}

We follow \cite{prentice1975discrimination}'s original parameterization, which is also used by \cite{cox2008generalized} for the PDF of the generalized F distribution:

\begin{eqnarray}
\label{pdf3}
f(t;m_1,m_2,\beta,\sigma) = \frac{1}{\sigma t B(m_1,m_2)} \frac{[\texttt{exp}(\frac{-\beta}{\sigma})(\frac{m_1}{m_2})t^{\frac{1}{\sigma}}]^{m_1}}{[1+\texttt{exp}(\frac{-\beta}{\sigma})(\frac{m_1}{m_2})t^{\frac{1}{\sigma}}]^{m_1+m_2}} 
\end{eqnarray}

where $m_1$,$m_2$,$\beta$, and $\sigma$ are parameters of generalized F distribution and $B$(,) is a beta function. 

We represent MRL using:

\begin{eqnarray}
\label{mrl20}
MRL(x) = \frac{\int_{x}^{\infty}tf(t)\texttt{dt}}{S(x)} - x
\end{eqnarray}

The numerator of equation \ref{mrl20} is what to focus on:

\begin{eqnarray}
\label{numerator_genf}
\int_{x}^{\infty}tf(t)dt &=& \int_{x}^{\infty}\frac{t}{\sigma t B(m_1,m_2)} \frac{[\texttt{exp}(\frac{-\beta}{\sigma})(\frac{m_1}{m_2})t^{\frac{1}{\sigma}}]^{m_1}}{[1+\texttt{exp}(\frac{-\beta}{\sigma})(\frac{m_1}{m_2})t^{\frac{1}{\sigma}}]^{m_1+m_2}}dt\\ 
\nonumber
&=& \frac{1}{\sigma B(m_1,m_2)} \int_{x}^{\infty}t \frac{\frac{1}{t}[\texttt{exp}(\frac{-\beta}{\sigma})(\frac{m_1}{m_2})t^{\frac{1}{\sigma}}]^{m_1}}{[1+\texttt{exp}(\frac{-\beta}{\sigma})(\frac{m_1}{m_2})t^{\frac{1}{\sigma}}]^{m_1+m_2}}dt
\end{eqnarray}

Using a substitution with $u$ = $\texttt{exp}(\frac{-\beta}{\sigma})(\frac{m_1}{m_2})t^{\frac{1}{\sigma}}$, thus $du = \frac{1}{\sigma t}\texttt{exp}(\frac{-\beta}{\sigma})(\frac{m_1}{m_2})t^{\frac{1}{\sigma}}dt$.
$u(x)$ = $\texttt{exp}(\frac{-\beta}{\sigma})(\frac{m_1}{m_2})x^{\frac{1}{\sigma}}$; $u(\infty) = \infty$. $t$ = $[\texttt{exp}(\frac{\beta}{\sigma})(\frac{m_2}{m_1})u]^{\sigma}$.

For the ease of exposition, we further let $C$ = $\texttt{exp}(\frac{-\beta}{\sigma})(\frac{m_1}{m_2})x^{\frac{1}{\sigma}}$.

\begin{eqnarray}
\label{simple10}
\int_{x}^{\infty}tf(t)dt &=& \frac{1}{B(m_1,m_2)} \int_{C}^{\infty}[\texttt{exp}(\frac{\beta}{\sigma})(\frac{m_2}{m_1})u]^{\sigma}\frac{u^{m_1-1}}{(1+u)^{m_1+m_2}}du\\
\nonumber
&=& \frac{\texttt{exp}(\beta)(\frac{m_2}{m_1})^{\sigma}}{B(m_1,m_2)} \int_{C}^{\infty}\frac{u^{m_1+\sigma-1}}{(1+u)^{m_1+m_2}}du
\end{eqnarray}

Using another substitution with $v$ = $(u - C)/C$, thus $dv = \frac{du}{C}$. $v(C) = 0$; $v(\infty) = \infty$.  $u = vC + C$.

\begin{eqnarray}
\label{simple11}
=\frac{\texttt{exp}(\beta)(\frac{m_2}{m_1})^{\sigma}C}{B(m_1,m_2)} \int_{0}^{\infty}(vC + C)^{m_1+\sigma-1}{(1+vC + C)^{-(m_1+m_2)}}dv
\end{eqnarray}

Factoring out $C$,

\begin{eqnarray}
\label{simple12}
=\frac{\texttt{exp}(\beta)(\frac{m_2}{m_1})^{\sigma}C^{-m_2+\sigma}}{B(m_1,m_2)} \int_{0}^{\infty}(v + 1)^{m_1+\sigma-1}{(1 + v + \frac{1}{C})^{-(m_1+m_2)}}dv
\end{eqnarray}

Using the $2F1$ hypergeometric function integral representation\footnote{https://functions.wolfram.com/HypergeometricFunctions/Hypergeometric2F1/07/01/01/0002/}:

\begin{eqnarray}
\label{simple13}
\nonumber
2F1(a,b,c;z) == \frac{\Gamma(c)}{\Gamma(b)\Gamma(c-b)}\int_{0}^{\infty}t^{-b+c-1}(t+1)^{a-c}(1+t-z)^{-a}dt; c > b > 0
\end{eqnarray}

\begin{eqnarray}
\label{simple14}
=\frac{\texttt{exp}(\beta)(\frac{m_2}{m_1})^{\sigma}C^{-m_2+\sigma}}{B(m_1,m_2)}
\frac{\Gamma(m_2-\sigma)}{\Gamma(m_2-\sigma+1)}
2F1(m_1+m_2,m_2-\sigma,m_2-\sigma+1;-\frac{1}{C})
\end{eqnarray}

When $m_2 < \sigma$, the MRL is undefined. Thus, we can further simplify $\frac{\Gamma(m_2-\sigma)}{\Gamma(m_2-\sigma+1)}$ as $\frac{1}{m_2 - \sigma}$.

Putting altogether, the $MRL(x)$ for Generalized F (i.e., \code{genf.orig}) is:

\begin{eqnarray}
\label{simple15}
\frac{\texttt{exp}(\beta)(\frac{m_2}{m_1})^{\sigma}}{B(m_1,m_2)}\frac{C^{-m_2+\sigma}}{m_2 - \sigma}
\frac{2F1(m_1+m_2,m_2-\sigma,m_2-\sigma+1;-\frac{1}{C})}{ S(x)} - x
\end{eqnarray}

While this expression is correct for any $C$>0 (which means $x$ > 0), it is not defined when $C$ = 0 or $x$ = 0 and turns MRL into the simple expectation.  However we can use the similar steps and derive the expectation of the Generalized F distribution directly, which is:

\begin{eqnarray}
\label{simple16}
E[T] = \texttt{exp}(\beta)(\frac{m_2}{m_1})^{\sigma}\frac{B(m_1+\sigma,m_2-\sigma)}{B(m_1,m_2)}
\end{eqnarray}

Again, when $m_2 < \sigma$, the expectation is undefined.  









\section{MRL Derivation for Prentice Generalized F}
In \cite{prentice1975discrimination}'s other  version of generalized F  distribution, $m_1$ and $m_2$ are replaced with shape parameters $Q$, and $P$, where $P > 0$ and:
$$m_1 = 2(Q^2 + 2P + Q\delta)^{-1} \quad \quad m_2 = 2(Q^2 + 2P - Q\delta)^{-1} \quad \quad \delta = (Q^2 + 2P)^{1/2}$$

We can use the following identity to equate the two parameterizations:
\begin{CodeChunk}
\begin{CodeInput}
 R> dgenf(x, mu=mu, sigma=sigma, Q=Q, P=P) 
 == dgenf.orig(x, mu=mu, sigma=sigma/ ((Q^2 + 2P)^{1/2}), 
 s1=2*(Q^2+2*P+Q*((Q^2 + 2P)^{1/2}))^(-1), 
 s2=2*(Q^2+2*P-Q*((Q^2 + 2P)^{1/2}))^(-1))

\end{CodeInput}
\end{CodeChunk}

\end{appendix}

\bibliography{refs}
\end{document}